\documentclass[prd, preprint, longbibliography, 12pt]{revtex4-1}

\usepackage{amsmath}
\usepackage{mathtools}
\usepackage{amssymb}
\usepackage{setspace}
\usepackage{graphicx}
\usepackage{natbib}
\usepackage{float}
\usepackage{tensor}
\usepackage[utf8]{inputenc}
\usepackage{amsfonts}
\usepackage{braket}
\usepackage{esint}
\usepackage{breqn}
\usepackage{IEEEtrantools}

\begin{document}
 \pagenumbering{gobble}
 %

\begin{center}
{ \large \bf Why do elementary particles have such strange  mass ratios?}\\
{\large - {\it The importance of quantum gravity at low energies}  \ - }

\bigskip

\bigskip

\vskip 0.1 in

{\large{\bf Tejinder P.  Singh }}

\smallskip

{\it Tata Institute of Fundamental Research,}
{\it Homi Bhabha Road, Mumbai 400005, India}\\
\smallskip
 {\tt tpsingh@tifr.res.in}

\end{center}
\vskip 1 in

\centerline{\bf ABSTRACT}
\smallskip

\noindent When gravity is quantum, the point structure of space-time should  be replaced by a non-commutative  geometry. This is true even for quantum gravity in the
infra-red. Using the octonions as space-time coordinates, we construct a pre-spacetime, pre-quantum Lagrangian dynamics. We show that the symmetries of this non-commutative space unify  the standard model of particle physics with $SU(2)_R$ chiral gravity. The algebra of the octonionic space  yields spinor states which can be identified with three generations of quarks and leptons. The geometry of the space implies quantisation of electric charge, and  leads to a theoretical derivation of the mysterious mass ratios of quarks and the charged leptons. Quantum gravity is quantisation not only of the gravitational field, but also of the point structure of space-time.

\vskip 1 in


\bigskip

\centerline{\it This article is an expanded version of an}
\centerline{Essay written for the Gravity Research Foundation 2022 Awards for Essays on Gravitation}
\centerline{This essay received an Honorable Mention}

\newpage

\pagenumbering{arabic}
\section{When is quantum gravity necessary?}
\noindent 
Consider a massive object in a quantum superposition of its two different classical position states $A$ and $B$. The resulting gravitational field is also then in a superposition, of the field corresponding to position $A$ and the field corresponding to position $B$. A clock kept at a field point $C$ will not register a definite value of time, nor a measurement of the metric will yield a well-defined result \cite{Penrose}. Let us now imagine a thought experiment in which every object in today's universe is in a superposition of its two different position states. The space-time metric will then undergo quantum fluctuations. Now, the Einstein hole argument shows that in order for space-time points to be operationally distinguishable,  the manifold must be overlaid by a (classical) metric \cite{Carlip}. Therefore, in our thought experiment, the point structure of space-time is lost, even though the energy scales of interest are much smaller than Planck scale, and the gravitational fields are weak.

When we describe microscopic systems by the laws of quantum theory, we take it for granted that the universe is dominated by classical bodies, so that a background space-time can be achieved and is available for defining time evolution of quantum systems. However, if everything were to be quantised at once, in the sense of the afore-mentioned thought experiment, no classical time will be available, and yet we ought to be able to describe the dynamics. This is an example of quantum gravity in the infra-red: the action of the gravitational field is much larger than $\hbar$ (unlike for Planck scale quantum gravity), and yet the point structure of space-time is lost. The manifold has to be replaced by something non-classical: quantum gravity is quantisation not only of the gravitational field, but also of the point structure of space-time.

Since the energy scale is not a relevant criterion for deciding whether gravity is classical or quantum, we propose that a gravitational field is quantum in nature when one or more of the following three (energy independent) criteria are satisfied: (i) the time scales of interest are of the order of Planck time $t_P$; (ii) the length scales of interest are of the order of Planck length $L_P$; and (iii) every sub-system has an action of the order $\hbar$ (and is hence quantum and obeys quantum superposition). If (iii) holds but (i) and (ii) do not, we have quantum gravity in the infra-red. If (iii) holds along with (i) and (ii) then we have quantum gravity in the UV. 

Put differently, there ought to exist a reformulation of quantum (field) theory, even at low energies,  which does not depend on classical time. Such a reformulation is essential  also for the standard model of particle physics. In fact we show that it helps us understand why the standard model has the symmetries it does, and why it's free parameters take the specific values they do, and also shows how to unify gravity with the other fundamental forces: electroweak and strong. We construct such a dynamics using Planck time $t_P$, Planck length $L_P$ and Planck's constant $\hbar$ as the only three fundamental parameters in the theory. We note that in these units the low energy fine structure constant $\alpha_f = e^2/\hbar c \equiv e^2\; t_P/\hbar L_P \sim 1/137$ is order unity and hence quantum gravitational in origin (QG in IR). On the other hand, particles masses $m \sim \epsilon m_P \equiv \epsilon \hbar t_P / L_P^2$ are not, because $\epsilon \ll 1$. However, mass ratios (at low energies) can be, and in fact are, quantum gravitational in origin.

To achieve our goal, we build on Adler's pre-quantum theory, i.e.  trace dynamics (TD) \cite{Adler1, Adler2}. Starting from classical Lagrangian dynamics, TD retains the classical space-time manifold, but all configuration variables and their canonical momenta are raised to the status of matrices (equivalently operators). This step is the same as in quantum theory; however the canonical Heisenberg commutation relations $[q,p]=i\hbar$ are not imposed. Instead, we have a matrix-valued Lagrangian dynamics, where the Lagrangian is the trace of a matrix polynomial made from matrix-valued configuration variables and their time derivatives (i.e. the velocities). A `trace' derivative enables the derivation of Lagrange equations of motion, and a global unitary invariance of the trace Hamiltonian (this being an elementary consequence of invariance of the trace under cyclic permutations) implies the existence of the novel conserved Noether charge $\widetilde{C} \equiv \sum_i \; [q_i, p_i]$. The Hamiltonian of the theory is in general not self-adjoint, and dynamical evolution is not restricted to be unitary. Assuming this dynamics to hold on Planck time scale resolution, one asks what the averaged dynamics on lower energy scales will be, if one coarse-grains the dynamics on time scales much larger than Planck time. Using the techniques of statistical thermodynamics it is shown that if the anti-self-adjoint part of the Hamiltonian is negligible, the emergent dynamics is relativistic quantum (field) theory. The afore-mentioned Noether charge is equi-partitioned over all bosonic and fermionic degrees of freedom, and canonical commutation and anti-commutation relations emerge for the statistically averaged canonical variables, which obey the Heisenberg equations of motion. If the anti-self-adjoint part of the Hamiltonian becomes significant (this is enabled by large-scale quantum entanglement), spontaneous localisation results, leading to the quantum-to-classical transition and emergence of classical dynamics. For a detailed explanation of the emergence of the classical universe the reader is referred to Section XIII of \cite{Singhreview}.

\section{Replacing the point structure of spacetime by the non-commutative geometry of the octonions}
\noindent Next, TD is  generalised, so as to replace the 4D Minkowski space-time manifold by a higher dimensional non-commutative space-time, and incorporate matrix-valued pre-gravitation, thus taking TD to a pre-space-time, pre-quantum theory. Let us recall that in special relativity, given the four-vector $V^\mu = dt \; \hat{t} + dx \;\hat{x} + dy \; \hat{y} + dz\; \hat{z}$   connecting two neighbouring space-time points having a separation $(dt, dx, dy, dz)$, one can define the line element $ds^2 = \eta_{\mu\nu} V^\mu V^\nu$ and the four-velocity $dq^\mu/ds$ of a particle having the configuration variable $q^\mu = (q^t, q^x, q^y, q^z)$. The action for the particle is $mc\int ds$ and the transition to curved space-time and general relativity is made by introducing the metric $g_{\mu\nu}$, i.e. $ds^2 = g_{\mu\nu} dx^{\mu} dx^{\nu}$, and writing down the action
\begin{equation} 
S = \frac{c^4}{16\pi G} \int d^4x\; \sqrt{-g}\; R \ +\ \sum_i \;m_i c \int ds \ + \ S_{YM}
\label{one}
\end{equation}
Here, the first term is the Einstein-Hilbert action, and $S_{YM}$ stands for the action of Yang-Mills fields, and also includes their current sources.

We now generalise this action to construct a pre-spacetime, pre-quantum action principle \cite{Roy} from which the sought for quantum theory without classical time emerges, and whose symmetries imply  the standard model of particle physics and fix its free parameters. The space-time coordinates $(t, x, y, z)$ are replaced by a set $\{e_i, \ i=0,1, 2,...,m-1\}$ of $m$ non-commuting coordinates,  to be specified later in this section. The configuration variable $q^\mu$ for a particle is replaced by a matrix $q_F$ whose entries are odd-grade Grassmann elements over the field of complex numbers (so as to represent fermions). $q_F$ has $m$ components $q_F^i$, one for each of the coordinates $e_i$, i.e. $q_F = (q_F^0\; e_0 + ....q_F^{(m-1)}\;e_{m-1})$. The point structure of space-time is lost; instead we have a non-commutative geometry, and the matrix-valued velocity is defined as $dq_F/d\tau \equiv \dot{q}_F$. Here, the newly introduced Connes time $\tau$ is a unique property of a non-commutative geometry; it is an absolute real-valued time parameter distinct from the non-commuting coordinates $e_i$, and is used to describe evolution \cite{Connes}.

To introduce pre-gravitation into trace dynamics, we recall the spectral action principle of Chamseddine and Connes, according to which the Einstein-Hilbert action can be cast in terms of the eigenvalues of the square of the (regularised) Dirac operator $D_B$  on a Riemannian manifold, by making use of a truncated heat kernel expansion \cite{Chamseddine}
\begin{equation}
Tr \; [L_P^2 \; D_B^2] \sim \int d^4x \sqrt{g} \; \frac{R}{L_P^2} + {\cal O}(L_P^0)\sim L_P^2 \sum_n \lambda_n^2
\label{grraction}
\end{equation}
Here, the eigenvalues $\lambda_n$ of the Dirac operator  play the role of  dynamical variables of general relativity \cite{Rovelli}.
Following trace dynamics, each eigenvalue  $\lambda_n$ is raised to the status of a canonical matrix momentum: $\lambda_n\rightarrow p_{Bn}\propto q_{Bn}/d\tau \equiv D_B$, and the bosonic matrix $q_B$ (with even grade Grassmann elements as entries) is now the  configuration variable, and it has $m$ matrix components $q_B^m$ over the non-commuting coordinates $e_i$.  Therefore we have $N$ copies of the Dirac operator ($n$ runs from 1 to $N$,  with  $N\rightarrow \infty$).  The trace Lagrangian [space-time part] of the matrix dynamics  for the $n$-th degree of freedom is given by $L_P^2\; Tr \; (dq_{Bn}/d\tau)^2$. The full action for the total matrix dynamics  [space-time part] is  $S \sim \sum_n \int d\tau\; L_P^2\; Tr\; (dq_{Bn}/d\tau)^2$. Yang-Mills fields are expressed by the matrices $q_{Bn}$, pre-gravitation by the $\dot{q}_{Bn}$,  the fermionic degrees of freedom by  fermionic matrices $q_{Fn}$ and by their `velocities' $\dot{q}_{Fn}$.
Each of the $n$ degrees of freedom has a fundamental action, which is given  by \cite{Singhreview, MPSingh}
\begin{equation}
\frac{S}{\hbar} =  \frac{a_0}{2} \int \frac{d\tau}{\tau_{Pl}} \; Tr  \bigg[\dot{q}_B^{\dagger} + i\frac{\alpha}{L} q_B^\dagger+ a_0 \beta_1\left( \dot{q}_F^\dagger  + i\frac{\alpha}{L} q_F^\dagger\right)\bigg] \times \bigg[ \dot{q}_B + i\frac{\alpha}{L} q_B+ a_0 \beta_2\left( \dot{q}_F + i\frac{\alpha}{L} q_F\right)\bigg] 
\label{ymi}
\end{equation} 
where  $a_0 \equiv L_P^2 / L^2$. The net action of this generalised trace dynamics is therefore the sum over $n$ of $N$ copies of the above action, one copy for each degree of freedom, and this new action replaces (\ref{one}) in the pre-theory. This full action defines the pre-spacetime, pre-quantum theory, with each degree of freedom [defined by the above action] considered as  an `atom' of space-time-matter [an STM atom].  $L$ is a length parameter [scaled with respect to $L_P$; $q_B$ and $q_F$ have dimensions of length] which characterise the STM atom, and $\alpha$ is the dimensionless Yang-Mills coupling constant. $\beta_1$ and $\beta_2$ are two unequal complex Grassmann numbers \cite{Singhreview}.

The subsequent analysis of this pre-space-time, pre-quantum theory is carried out analogously to the pre-quantum trace dynamics. Equations of motion are derived, and there is again a conserved Noether charge.  Assuming that the theory is valid at the Planck time scale, the coarse-grained emergent low-energy approximation obeys quantum commutation rules and Heisenberg equations of motion, and  this is also the sought for reformulation of quantum theory without classical time. The emergent dynamics is also the desired quantum theory of gravity in the infra-red. If a sufficient number of STM atoms get entangled, the anti-self-adjoint part of the Hamiltonian becomes important, and  spontaneous localisation results; the fermionic part of the entangled STM atoms gets localised. There hence emerges a 4D classical space-time manifold (labelled by the positions of collapsed fermions), which is sourced by point masses and by gauge fields, and  whose geometry obeys the laws of general relativity given above by (\ref{one}). Those STM atoms which are not sufficiently entangled continue to remain quantum; their dynamics is described by the low energy pre-theory itself, or approximately by quantum field theory on  the 4D space-time background generated by the entangled and collapsed fermions [these being the macroscopic bodies of the universe]. 

Note that the non-commutative coordinate system $\{e_i\; ;  \ i=1,2,...,n\}$ is not impacted by the coarse-graining. The averaging takes place only over the time-scale $\tau$ and hence over energy; therefore the non-commuting coordinates $e_i$ remain valid at low energies as well. What then, should we choose as our $e_i$, in place of the four real numbers $(t, x, y, z)$ which label the 4D space-time manifold in classical physics? We take clue from the normed division algebras, i.e. number systems in which the four operations of addition, subtraction, multiplication and division can be defined. There are only four such number systems: real numbers $\mathbb R$, complex numbers $\mathbb C$, quaternions $\mathbb H$, and the octonions $\mathbb O$. A quaternion $H = (a_0 e_0 + a_1 \hat{i} + a_2  \hat{j} + a_3 \hat{k})$ is a generalisation of complex numbers, such that the $a_i$ here are reals, $e_0^2=1$ and
\begin{equation}
\hat{i}^2 = \hat{j}^2 = \hat{k}^2 = -1\; ; \quad \hat{i} \hat{j} = - \hat{j} \hat{i} = \hat{k} ; \quad \hat{j} \hat{k} = - \hat{k} \hat{j} = \hat{i}; \quad \hat{k} \hat{i} = - \hat{i} \hat{k} = \hat{j}
\end{equation}
Quaternions are used to describe rotations in three dimensions, and the automorphism group formed by the three imaginary directions is $SO(3)$. Complex quaternions $\mathbb C \times \mathbb H$ generate the Lorentz algebra $SL(2,\mathbb C)\sim SO(1,3)$ and the Clifford algebra $Cl(2)$, if one of the three quaternionic imaginary directions is kept fixed. If no direction is kept fixed, they generate the Lorentz algebra in 6D : $SL(2, \mathbb H)\sim SO(1,5)$ and the Clifford algebra $Cl(3)$.  However, they are not a big enough number system for unifying all the standard model symmetries  with the Lorentz symmetry. Whereas, the octonions seem to be just right for that purpose!

An octonion is defined as $O = a_0 e_0 + a_1 e_1 + a_2 e_2 + a_3 e_3 + a_4 e_4 + a_5 e_5 + a_6 e_6 + a_7 e_7$ such that the $a_i$ are reals, $e_0^2=1$, each of the seven imaginary directions $(e_1, e_2, ..., e_7)$ squares to $-1$, these directions anti-commute with each other, and their multiplication rule is given by the so-called Fano plane. Octonionic multiplication is non-associative. The imaginary directions form the automorphism group $G_2$, which is the smallest of the five exceptional Lie groups $G_2, F_4, E_6, E_7, E_8$ all of which have to do with the symmetries of the octonion algebra. $F_4$ is the automorphism group of the exceptional Jordan algebra: the algebra of $3\times 3$ Hermitean matrices with octonionic entries, and $E_6$ is the automorphism group of the complexified exceptional Jordan algebra \cite{Ramond}. The octonions are our sought for non-commuting coordinates $e_i$ on which the action principle (\ref{ymi}) is constructed.  They generate 10D space-time : $SL(2,\mathbb O)\sim SO(1,9)$. The coordinate geometry of the octonions dictates the allowed symmetry groups, and definition and properties of  fermions such as quantisation of electric charge \cite{Furey}, value of the low energy fine structure constant \cite{Singh}, and mass-ratios \cite{Bhatt}. The parameter $L$ and the coupling constant $\alpha$ in (\ref{ymi}) are determined by the algebra of the octonions, not by the dynamics of $q_F$ and $q_B$. This way, not only does the geometry tell matter how to move, it also tells matter what to be. The dynamical variables $(q_B, q_F)$ curve the flat geometry $\{e_i\}$; however even before the dynamics is switched on, the low-energy standard model of particle physics is fixed by the $e_i$, unlike when space-time is $\mathbb R^4$. The transition $e_i\rightarrow q_F^i e_i + q_B^ie_i$ is akin to the transition $\eta_{\mu\nu}x^\mu x^\nu\rightarrow g_{\mu\nu} x^\mu x^\nu$, with the important difference that the former transition takes place at the `square-root of metric' level, as if for tetrads, and the matrices $q_B$ and $q_F$ incorporate standard model forces besides gravity, and also fermionic matter. In fact, with the redefinition
${\dot{\widetilde{Q}}_B} \equiv (i\alpha q_B + L \dot{q}_B)/L; \; {\dot{\widetilde{Q}}_F} \equiv  (i\alpha q_F + L \dot{q}_F)/L$
the Lagrangian in (\ref{ymi}) can be brought to the elegant and revealing form, as if describing  a two-dimensional  (because $\beta_1\neq \beta_2$) free particle:
\begin{equation}
 \frac{S}{\hbar} =  \frac{a_0}{2} \int \frac{d\tau}{\tau_{Pl}} \;  Tr  \biggl(\biggr. \dot{\widetilde{Q}}_{B}^\dagger + \dfrac{L_{p}^{2}}{L^{2}} \beta_{1} \dot{\widetilde{Q}}_{F}^{\dagger} \biggl.\biggr) \biggl(\biggr. \dot{\widetilde{Q}}_{B} + \dfrac{L_{p}^{2}}{L^{2}} \beta_{2} \dot{\widetilde{Q}}_{F} \biggl.\biggr) \biggl.
\label{eq:tracelagn}
\end{equation}
In this fundamental form of the action, the coupling constant $\alpha$ is not present. In fact $\alpha$, along with mass ratios, emerges only after (left-right) symmetry breaking segregates the unifying dynamical variable $\dot{\widetilde{Q}}_B$ into its gravitational part $\dot{q}_B$ and Yang-Mills part $q_B$. 

\section{Spinor states for quarks and leptons, from  the algebra of the complex octonions}
\noindent The automorphism group $G_2$ of the octonions has two maximal sub-groups $SU(3)$ and $SO(4)\sim SU(2) \times SU(2)$, the first of which is the element preserver group of the octonions, and the second is the stabiliser group of the quaternions inside the octonions \cite{Todorov}. The two groups have a $SU(2)$ intersection. Keeping one of the seven imaginary directions, say $e_7$, fixed, the remaining six directions can be used to form an MTIS (maximally totally isotropic subspace) and the following generators (along with their adjoints) for the Clifford algebra $Cl(6)$:
\begin{equation}
    \alpha_1 = \frac{-e_5 + ie_4}{2}, \quad \alpha_2 = \frac{-e_3 + ie_1}{2}, \quad \alpha_3 = \frac{-e_6 + ie_2}{2}
\end{equation}
(This is a covariant choice as all the imaginary directions are equivalent and interchanging any of them does not change the analysis or results). From here, one can construct spinors as minimum left ideals of the algebra, by first constructing the idempotent  $\Omega\Omega^{\dagger}$ where $\Omega = \alpha_1\alpha_2\alpha_3$.  The eight resulting spinors are 
\begin{equation}
\begin{split}
& {\mathcal{V}} = \Omega\Omega^{\dagger} ; \quad
    V_{ad1} = \alpha_{1}^{\dagger}\mathcal{V} ; \quad
    V_{ad2} = \alpha_{2}^{\dagger}\mathcal{V} ; \quad
    V_{ad3} = \alpha_{3}^{\dagger}\mathcal{V} ;   \\
   &  V_{u1} = \alpha_{3}^{\dagger}\alpha_{2}^{\dagger}\mathcal{V} ;\quad
    V_{u2} = \alpha_{1}^{\dagger}\alpha_{3}^{\dagger}\mathcal{V}; \quad
    V_{u3} = \alpha_{2}^{\dagger}\alpha_{1}^{\dagger}\mathcal{V} ; \quad
    V_{e+} = \alpha_{3}^{\dagger}\alpha_{2}^{\dagger}\alpha_{1}^{\dagger}\mathcal{V} 
    \end{split}
    \label{spinfem}
\end{equation} 
After defining the operator $Q = (\alpha_1^{\dagger}\alpha_1 + \alpha_2^{\dagger}\alpha_2 + \alpha_3^{\dagger}\alpha_3)/3$ as one-third of the $U(1)$ number operator we find that the states $\mathcal{V}$ and $V_{e+}$ are singlets under $SU(3)$ and respectively have the eigenvalues $Q=0$ and $Q=1$. The states $V_{ad1}, V_{ad2}, V_{ad3}$ are anti-triplets under $SU(3)$ and have $Q=1/3$ each, whereas the states $V_{u1}, V_{u2}, V_{u3}$ are triplets under $SU(3)$ and each have $Q=2/3$. These results allow $Q$ to be interpreted as electric charge, and the eight states represent a neutrino, three anti-down quarks, three up quarks and the positron having the standard model symmetries $SU(3)_{color}\times U(1)_{em}$. Anti-particle states are obtained by complex conjugation. The eight $SU(3)$ generators can also be expressed in terms of the octonions and represent the eight gluons, whereas the $U(1)$ generator is for the photon. We hence see the standard model of particle physics emerging from the symmetries of the physical octonionic space, and the quantisation of electric charge is a consequence of the coordinate geometry of the octonions \cite{Furey}.

To see how the weak force (and electroweak) and chiral gravity emerge from the other maximal sub-group $SO(4) \sim SU(2)\times SU(2)$ we must consider three fermion generations and the larger exceptional Lie group $E_6$ because these symmetries are shared pair-wise across fermion generations, as shown in Fig. 1. Furthermore, the neutrino will be assumed to be Majorana, because only then the correct values of mass ratios are obtained \cite{Bhatt}. Also, notably, $E_6$ is the only one of the exceptional groups which has complex representations.

\section{Pre-gravitation and mass ratios from a left-right symmetric extension of the standard model of particle physics}
\noindent The 78 dimensional exceptional Lie group $E_6$ is the automorphism group of the complexified Jordan algebra, and admits the sub-group structure shown in Fig. 1, as motivated by the discussion in \cite{DrayE6}. $E_6$ contains three intersecting copies of $Spin(9,1)\sim SL(2,\mathbb O)$ which have an $SO(8)$ intersection, and the triality property of $SO(8)$ motivates that there are exactly three fermion generations. In order to account for the symmetries of $E_6$ and to obtain chiral fermions, we now work with split bioctonions (instead of octonions, which are used in (\ref{eq:tracelagn}) i.e. before symmetry breaking) \cite{Vatsalya}. 
\begin{figure}[h]
\centering
\includegraphics[width=13cm]{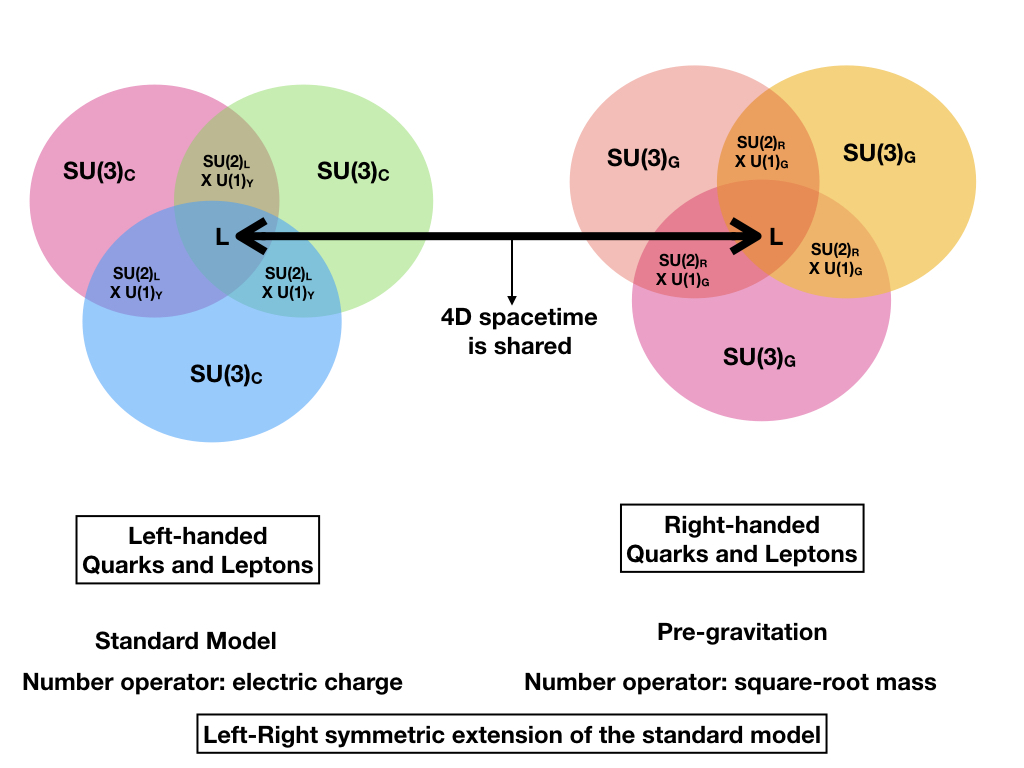}
\caption{Unification from symmetries of $E_6$}
\end{figure}
Embedded in the three $Spin(9,1)$ are three copies of $SU(3)$, one of which is $SU(3)_{color}$ and one is for generational symmetry - this is shown in the left part of Fig. 1. There is a pairwise intersection amongst a pair of generations, which is the electroweak group $SU(2)_L\times U(1)_Y$ from which the weak interaction and electromagnetism can be obtained. There is a three way intersection marked $L$ which is the 4D Lorentz group $SO(3,1)$. The $Cl(4)$ generators of $SU(2)_L$ are made from the $Cl(6)$ of $SU(3)_{color}$ and $Cl(2)$ of the Lorentz algebra, and it can be shown that $SU(2)_L$ acts only on left-handed fermions. The spinor states for the LH quarks and leptons of one generation are constructed analogously to those in (\ref{spinfem}), by using the left-handed active Majorana neutrino as the idempotent, with complex conjugation giving the corresponding antiparticles. The spinor states for the second and third generation are respectively obtained by applying two successive $2\pi/3$ rotations on the eight states of the first generation while staying in the plane defined by the form $(e_i + i e_j)$ of a given first generation particle [SO(8) symmetry implies eight independent great circles on an 8-sphere, one for each of the eight particles, and three particles of three generations]. There are  a total of 24 LH fermions and their 24 anti-particles, and 12 gauge bosons. The unified symmetry group of  Lagrangian (\ref{eq:tracelagn})  is $E_6$.

Similarly, three generations of RH fermions are obtained by using split octonions and the three RH sterile Majorana neutrinos as the idempotent. We identify the associated $SU(3)$ with $SU(3)_{grav}$ - a newly introduced gravitational sector; and identify the $U(1)$ number operator with [$\pm$ square-root] of the mass of a quark / lepton (in Planck mass units), and the eight respective spinor states of one generation are: the sterile neutrino, three positrons of three different gravi-colors, three RH up quarks of three colors same as $SU(3)_{color}$, and one down quark which is a singlet under $SU(3)_{grav}$, along with the singlet sterile neutrino \cite{Priyank}. These obtain the respective square-root mass number $(0, 1/3, 2/3, 1)$ explaining why the down quark is nine times heavier than the electron. The $SU(2)_R$ is RH chiral gravity (LQG?) \cite{Ashtekar} which  reduces, in the classical limit induced by spontaneous localisation, to general relativity. The Lorentz group (whose Casimir invariant is the introduced mass number) is common with the LH particles, and its 6 dimensions, together with the 24 RH fermions and 12 new gauge bosons, when added to the LH sector, give the correct count of 78 for $E_6$. The split complex number gives  a scalar field which acts as the Higgs mediating between the LH charge eigenstates, and RH mass eigenstates. Is $U(1)$ gravity the  dark energy? This possibility is discussed further in Section V.

The group $E_6$ is also the symmetry group for the Dirac equation in 10D \cite{DrayE6} for three fermion generations (either LH or RH). The eigenvalue and eigenmatrix problem for the Dirac equation is in fact the same as $ J_3 (8) X = \lambda X$ where $J_3(8)$ is the exceptional Jordan algebra with symmetry group $F_4$. Substituting the above-mentioned spinor states of LH fermions (these being eigenstates of electric charge) and solving this eigenvalue problem expresses the LH charge eigenstates as superpositions of the RH mass eigenstates (thus fixing $\alpha$ and $L$ in (\ref{ymi})), and the ratios of the eigenvalues yield mass ratios of charged fermions as shown in Fig. 2; these exhibit very good agreement with the mysterious mass ratios, as shown in the table below \cite{Bhatt}.

$E_6$ has three copies of 10D spacetime. We never compactify the extra six complex dimensions - these represent the standard model  internal forces which determine the geometry of these extra dimensions. Quantum systems do not live in 4D spacetime. They live in $E_6$ and their true dynamics is the generalised trace dynamics, with evolution given by Connes time. Only classical systems live in 4D spacetime, where they descend as a result of spontaneous localisation of highly entangled fermions (compactification without compactification).  This overcomes the troublesome non-unique compactification problem of string theory.
\begin{figure}[h]
\centering
\includegraphics[width=15cm]{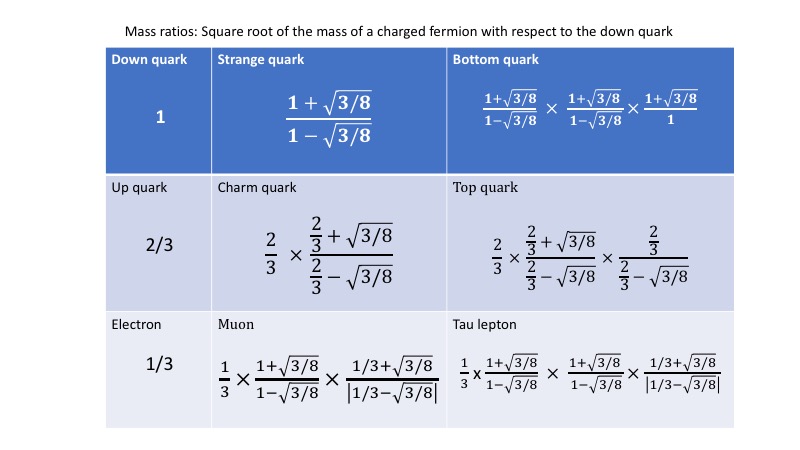}
\caption{Square-root mass ratios of charged elementary fermions \cite{Bhatt}}
\end{figure}

\begin{center}
\begin{tabular}{ |p{3cm}||p{3cm}|p{3cm}|p{3cm}|  }
 \hline
 \multicolumn{4}{|c|}{Square root mass ratios} \\
 \hline
 Particles & Theoretical mass ratio & Minimum experimental value& Maximum experimental value\\
 \hline
 muon/electron   & 14.10    &  14.37913078& 14.37913090 \\
 taun/electron &   58.64  & 58.9660   &58.9700\\
 charm/up & 23.57& 21.04 &  26.87\\
 top/up &289.26 & 248.18&  310.07\\
 strange/down &   4.16  & 4.21 & 4.86\\
 bottom/down & 28.44  & 28.25   &30.97\\
 
 \hline
 \end{tabular}
\end{center}
\centerline{Table I: Comparison of theoretically predicted square-root mass ratio with experimentally known range \cite{Bhatt}}
\bigskip

Apart from the two mass ratios of charged leptons, other theoretical mass ratios lie within the experimental bounds \cite{Zyla1}. On accounting for the so-called Karolyhazy correction \cite{Tp3}  we might possibly get more accurate mass ratios for all particles including charged leptons.  This will be investigated in future work.

In quantum theory, even at low energies, assuming a point structure for spacetime is an approximation; it is because of this approximation that the standard model of particle physics has so many unexplained free parameters. When we replace this approximate description  by a non-commutative spacetime, we find evidence that these parameter values get fixed. In particular, we derive the low energy fine structure constant \cite{Singh, Tp3} and mass ratios of charged fermions \cite{Bhatt} from first principles. We do not need experiments at ever higher energies to understand the low energy standard model. Instead, we need a better understanding of the quantum nature of spacetime at low energies, such that the quantum spacetime is consistent with the principle of quantum linear superposition.

\section{Further developments, clarifying remarks, and current status of the present unification programme}
The aforesaid essay is intended to give the reader a short overview of a new approach to quantum gravity and unification, details of which can be found in \cite{Singhreview}. In the present section we report on a few new insights not described in our earlier work, and provide clarifying details on some of the statements in the previous sections. 

One further way to motivate the present theory is to recall that when one takes the square root of the Klein-Gordon equation to arrive at the Dirac equation for spin-half fermions described by spinors, one does not take the corresponding square root of the four dimensional Minkowski spacetime labelled by four real numbers. But suppose one were to take the latter square root as well; then one arrives at a spinor description of spacetime, i.e. Penrose's twistor space, labelled by complex numbers, and one notes that $SL(2,C)$ is the double cover of the Lorentz group $SO(3,1)$. We can now try now to describe fermions on this twistor space [both the fermions as well as the space-time are now spinorial] and we can write down the Dirac equation on this twistor space. We can go even further and replace the complex numbers by one or the other of the only two additional division algebras, the quaternions ($H$) and the octonions $(O)$. Quaternionic twistor space is equivalent to 6D spacetime, because $SL(2,H)$ is the double cover of $SO(5,1)$, whereas octonion-valued twistor space is equivalent to 10D spacetime because $SL(2,O)$ is the double cover of $SO(9,1)$. When we describe fermions on octonionic twistor space we begin to deduce remarkable results such as three generations of quarks and leptons, and quantisation of electric charge, just as in the standard model. We are now solving the Dirac equation on the spinor equivalent of 10D Minkowski space-time, and this implies mass quantisation, a derivation of the low energy fine structure constant and of mass ratios of quarks and charged leptons. The bosonic sector now includes pre-gravitation in unification with the standard model forces, and the theory is shown to obey an $E_8 \times E_8$ symmetry \cite{Priyank}. Indeed, the theory we arrive at is a revised string theory without the troublesome non-uniqueness problem of compactification, and the fundamental entities are 2-branes with an area of the order of Planck area.

It is important to note that in going from real numbers to complex numbers to octonions, we have not changed the energy scale of the problem; rather we have gone from 4D Minkowski spacetime to 8D twistor space. In so doing we have simply found a new mathematical description of the standard model which explains its origin and its unification with pre-gravitation. We therefore arrive at the inescapable conclusion that elementary particles live in a space with $E_8 \times E_8$ symmetry even at low energies (built on a 10D complex non-commutative spacetime). They do not live in ordinary spacetime, and definitely not in 4D classical space-time. The extra dimensions are not Planck size, but  large extra dimensions, and have an absolute modulus of the order of the scale of the strong force and the weak force [$10^{-18}$ m to $10^{-15}$ m]. This can be approximately justified by recalling that in this theory the absolute modulus $l$ of the extra dimensions goes as $l\sim R^{1/3}$ (because of the holographic principle \cite{Singhholo}) where $R$ is the cosmic length scale, and both $l$ and $R$ are expressed in units of Planck length. Further work remains to be done to make this result rigorous, and also to derive the Higgs mass and the masses of the $W$ bosons from fundamental considerations.

The action principle in Eqn. (\ref{ymi}) is motivated by the action principle for a free particle in Newton's mechanics, which of course is nothing but the integral of the kinetic energy over absolute time. Now, over octonionic space [more precisely, complex split bioctonionic space] we have written the equivalent expression for the kinetic energy of a 2-brane over Connes time. The undotted variables are related to the left-chiral sector, this being the gauge bosons of the standard model and three generations of left chiral fermions, defined over octonionic space. The dotted variables are related to the right chiral sector, this being pre-gravitation $SU(3)_{grav}\times SU(2)_R \times U(1)_{grav}$ and three generations of right chiral fermions defined over the split part of bioctonionic space \cite{Priyank}. A detailed investigation of the Lagrangian as regards its particle content is currently in progress. $\dot{q}_B$ is the Dirac operator on octonionic space and $q_B$, the Yang-Mills field, is the correction to the Dirac operator, as in conventional quantum field theory. The spectral action principle tells us the classical limit of the trace of the Dirac operator squared, when Yang-Mills fields are present. This classical limit has been discussed briefly in our earlier work \cite{MPSingh} and is given by the following equation, from \cite{Chams} where $D_{Bnew} \equiv D_B + \alpha A$  is the corrected Dirac operator resulting after including the Yang-Mills potential $A$:
\begin{align*}
Tr\ [L_P^2\ D_{Bnew}^2 ] &= \dfrac{N}{48 \pi^{2}} \biggl[\biggr. 12 L_P^{-4} f_{0} \int d^{4}x\sqrt{g} + L_{P}^{-2} f_{2} \int d^{4}x\sqrt{g} R\\
&+ f_{4} \int d^{4}x\sqrt{g} \biggl[\biggr. -\dfrac{3}{20} C_{\mu \nu \rho \sigma} C^{\mu \nu \rho \sigma} + \dfrac{11}{20} R^{*} R^{*} + \dfrac{1}{10} R;_{\mu}\, ^{\mu}\\
&+ \dfrac{g^{2}}{N} F^{i}_{\mu\nu} F^{\mu\nu i} \biggl.\biggr] + \mathcal{O}\biggl(\biggr. {L_{P}^{2}} \biggl.\biggr) \biggl.\biggr]
\label{diremerge}
\end{align*}
where
\begin{itemize}
\item $\dfrac{N L_{P}^{-2} f_{2}}{48 \pi^{2}} \int d^{4}x\sqrt{g} R$ term is the Einstein-Hilbert action
\item $\dfrac{N L_{P}^{-4} f_{0}}{4 \pi^{2}} \int d^{4}x\sqrt{g}$ term is responsible for the cosmological constant
\item $\dfrac{f_{4} g^{2}}{48 \pi^{^{2}}} \int d^{4}x\sqrt{g} F^{i}_{\mu\nu} F^{\mu\nu i}$ term is the Yang-Mills action
\item $-\dfrac{N f_{4}}{320 \pi^{^{2}}} \int d^{4}x\sqrt{g} C_{\mu \nu \rho \sigma} C^{\mu \nu \rho \sigma}$ term would be responsible for the Conformal gravity
\item $\dfrac{11 N f_{4}}{960 \pi^{^{2}}} \int d^{4}x\sqrt{g} R^{*} R^{*}$ term would be responsible for the Gauss-Bonnet gravity
\end{itemize}
This is the expansion of the squared Dirac operator when gauge fields are included alongside gravity. We do not yet take into account the volume term, and  conformal gravity, and Gauss-Bonnet gravity in our present work. It is however an issue of great significance that in the classical limit, general relativity is being modified by conformal gravity, and encourages us to relate our theory to MOND and RMOND, as an alternative to cold dark matter. Interestingly, there so far seems to be no cold dark matter candidate particle in our theory, and MOND and sterile neutrinos seem to arise naturally in our approach to unification; please see section XIV of  \cite{Singhreview}; further work is currently in progress in this direction.

The matrix-valued equations of motion are easily written down after first defining
\begin{equation}
q_1 = q_B + a_0 \beta_1 q_F , \qquad q_2 = q_B + a_0 \beta_2 q_F
\end{equation}
where we have set $a_0 \equiv L_P^2 / L^2$.
In terms of these two variables, the above trace Lagrangian  can be written as
\begin{equation}
Tr {\cal L} = \frac{1}{2} a_1 a_0  Tr \left[\dot{q}_1  \dot{q}_2  - \frac{\alpha^2 c^2}{L^2} q_1 q_2 + i\frac{\alpha c}{L} \left(\dot{q}_1 q_2 + q_1 \dot{q}_2 \right) \right]
\label{lagnew}
\end{equation}
where $a_1 \equiv \hbar /cL_P $. The last term in the trace Lagrangian is a total time derivative, and hence does not contribute to the equations of motion, so that we can get dynamics from the Lagrangian:
\begin{equation}
Tr {\cal L} = \frac{1}{2} a_1 a_0  Tr \left[\dot{q}_1  \dot{q}_2  - \frac{\alpha^2 c^2}{L^2} q_1 q_2  \right]
\label{q1q2lag}
\end{equation}

The canonical momenta are given by
\begin{equation}
p_1 = \frac{\delta Tr {\cal  L}} {\delta \dot{q}_1}  = \frac{1}{2} a_1 a_0 \dot{q}_2 ; \qquad  p_2 = \frac{\delta Tr {\cal  L}} {\delta \dot{q}_2}  = \frac{1}{2} a_1 a_0 \dot{q}_1
\end{equation}
The Euler-Lagrange equations of motion are
\begin{equation}
\ddot{q}_1 = - \frac{\alpha^2 c^2}{L^2} q_1; \qquad \ddot{q}_2 = - \frac{\alpha^2 c^2}{L^2} q_2
\end{equation}
In terms of these two complex variables, the 2-brane behaves like two independent complex-valued oscillators. However, the degrees of freedom of the 2-brane couple with each other when expressed in terms of the self-adjoint variables $q_B$ and $q_F$. This is because $q_1$ and $q_2$ both depend on $q_B$ and $q_F$, the difference being that $q_1$ depends on $\beta_1$ and $q_2$ depends on $\beta_2$.

The trace Hamiltonian  is 
\begin{equation}
Tr{\cal H} = Tr [p_1 \dot{q}_1 + p_2 \dot{q}_2 - Tr {\cal  L}] = \frac{a_1 a_0}{2}Tr \left[ \frac{4}{a^2_1 a^2_0}p_1 p_2 + \frac{\alpha^2 c^2}{L^2} q_1  q_2 \right]
\end{equation}
and Hamilton's equations of motion are
\begin{equation}
\dot{q}_1  = \frac{2}{a_1 a_0} p_2  \qquad  \dot {q}_2  = \frac{2}{ a_1 a_0 } p_1  ;  \qquad
\dot{p}_1 = -\frac{a_1 a_0 \alpha^2  c^2}{2L^2} q_2; \qquad \dot{p}_2 = -\frac{a_1 a_0 \alpha^2  c^2}{2L^2} q_1
\end{equation}
It is understood that this generalised trace dynamics is defined over complex bioctonionic space, and elementary particles and gauge bosons including those for pre-gravitation are special cases of these dynamical variables, reminiscent of the different vibrations of the string in string theory.

\noindent {\bf Octonions, elementary particle physics, and gravitation:} Octonions have been associated with physics for a very long time. In fact already in the 1920s Jordan discovered the algebraic formulation of quantum mechanics, and the Jordan algebras, and in particular his work led to the discovery of the exceptional Jordan algebra $J_3(8)$ (also known as the Albert algebra). This is the algebra of 3x3 Hermitean matrices with octonionic entries. Since the 1970s there have been extensive investigations of how the octonions bear a fundamental connection with elementary particles, quarks as well as leptons of the standard model \cite{o1, o2, o3, o4, o5, o6, o7, o8, o9, o10, o11, o12,o13}.

How does our work compare with and differ from these earlier works on applications of octonions to quantum mechanics and particle physics? Our work is strongly inspired by these earlier investigations and builds on them, especially with regard to the definition of elementary particle states using Clifford algebras made from octonionic chains, applications of the exceptional Lie groups and the exceptional Jordan algebra. However, our fundamental perspective is different from these earlier works which largely focus on relating octonions to the standard model and its extensions such as GUTs, to the octonions. Our starting point is wholly quantum foundational: to seek a reformulation of quantum theory which does not depend on classical time. This turns out to be a pre-quantum, pre-spacetime theory which is a matrix-valued Lagrangian dynamics on an octonionic twistor space. Remarkably, it is also a theory of quantum (pre-)gravity and a unification of the standard model of particle physics with pre-gravitation. The eigenvalue problem for the Dirac equation on octonionic space [equivalently 10D spacetime] is solved by the characteristic equation of the exceptional Jordan algebra. In combination with the Lagrangian of the theory, this leads to a derivation of the low energy fine structure constant, and mass ratios of quarks and charged leptons. Thus, by bringing in gravitation, and trace dynamics, our work significantly expands the scope of earlier related research. This new  scenario is summarised in Fig. 3. Below, we elaborate on some aspects of these new developments.
\begin{figure}[h]
\centering
\includegraphics[width=15cm]{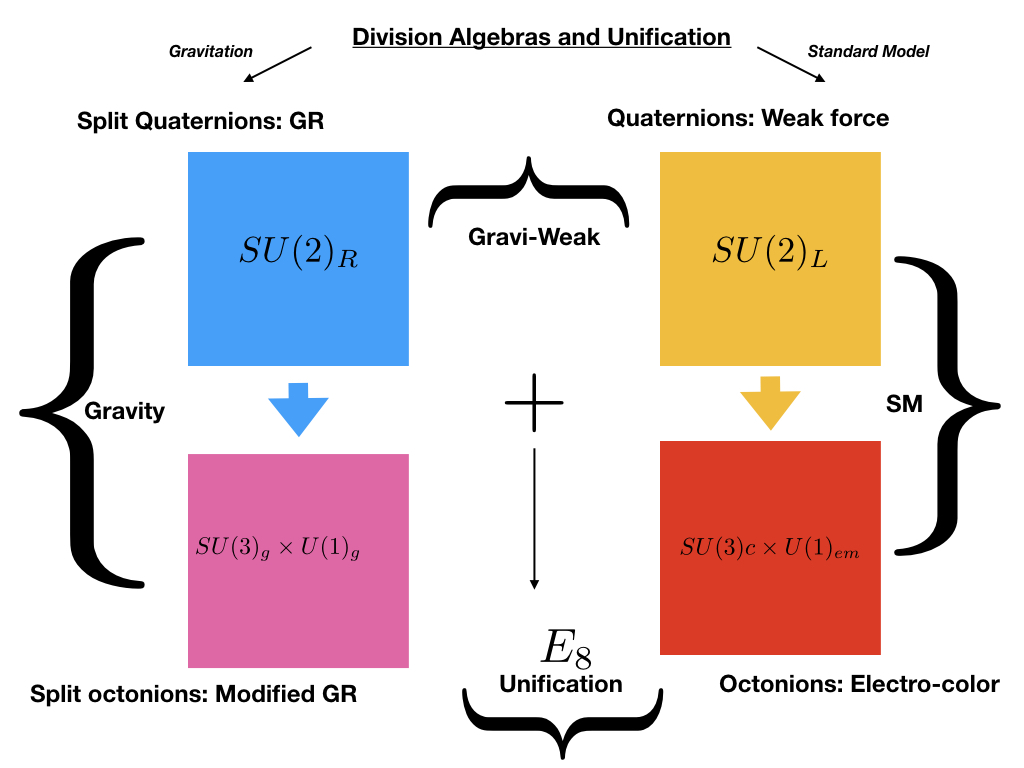}
\caption{Octonions, and an $E_8\times E_8$ unification of standard model and pre-gravitation \cite{Priyank}}
\end{figure}

\noindent {\bf Mass quantisation from a number operator:}
The masses of the electron, the up quark, and the down quark, are in the ratio 1 : 4 : 9
This simple fact calls for a theoretical explanation.
A few years back Cohl Furey proved the quantisation of electric charge as a consequence of constructing the states for quarks and leptons from the algebra of the octonions [arXiv:1603.04078 Charge quantisation from a number operator]. The complex  octonions are used to construct a Clifford algebra Cl(6) which is then  used to make states for one generation of quarks and leptons. The automorphism group $G_2$ of the octonions has a sub-group SU(3) and these particle states have the correct transformation properties as expected if this SU(3) is $SU(3)_{color}$ of QCD. Further, (one-third of) a number operator made from the Cl(6) generators has the eigenvalues (0, 1/3, 2/3, 1) [with 0 and 1 for the SU(3) singlets and 1/3, 2/3 for the triplets] allowing this to be identified with electric charge. This proves charge quantisation and the U(1) symmetry of the number operator is identified with $U(1)_{em}$. Anti-particle states obtained by complex conjugation of particle states are shown to have electric charge (0, -1/3, -2/3, -1). Thus the algebra describes the electro-colour symmetry for the neutrino, down quark, up quark,  electron, and their anti-particles. Note that it could instead be the second fermion generation, or the third generation. Each generation has the same charge ratio (0, 1/3, 2/3, 1).

This same analysis can now be used to show that the square-root of the masses of electron, up and down are in ratio 1:2:3 All we have to do is to identify the eigenvalues of the number operator with the square-root of the mass of an elementary particle, instead of its electric charge. And we also get a classification of matter and anti-matter, after noting that complex conjugation now sends matter to anti-matter, as follows:
\begin{align*}
& {\rm Matter} \quad\sqrt{mass}        \qquad     \qquad     \qquad       {\rm Anti-matter} \quad \sqrt{mass} \\
& Anti- Neutrino\   0              \qquad        \qquad \qquad        \qquad          Neutrino\ 0\\
& Electron\ 1/3     \qquad        \qquad \qquad        \qquad                                                positron \  -1/3\\
& Up-quark\ 2/3               \qquad        \qquad \qquad        \qquad                                    anti-up \ -2/3\\
& Down-quark \ 1        \qquad        \qquad \qquad        \qquad                                                      anti-down  \ -1
\end{align*}
Compared to the electric charge case above, the electron and down quark have switched places, and we already have our answer to the mass quantisation question asked at the start of this sub-section. There is again an SU(3) and a U(1) but obviously this is no longer QCD and EM. We identify this symmetry with a newly proposed $SU(3)_{grav} \times U(1)_{grav}$ whose full physical implications remain to be unravelled. [GR iemerges from $SU(2)_R$ this being an  analog of the weak force $SU(2)_L$].
$E_6 \times E_6$  admits a sub-group structure with two copies each of SU(3), SU(2) and U(1). Therefore, one set is identified with the standard model $SU(3) \times SU(3) \times U(1)$ (electric charge based) and the other with the newly introduced $SU(3)_{grav} \times SU(2)_R \times U(1)_{grav}$ (sqaure-root mass based). In the early universe, the separation of matter from anti-matter is the separation of particles with positive square-root mass from particles of negative square root  mass. This separation effectively converts the vector-interaction of pre-gravitation into an attractive only emergent gravitation.

However, the second and third fermion generations do not have the simple mass ratios (1, 4, 9) unlike the electric charge ratios which are same for all three generations. Why so?! Because mass eigenstates are not the same as charge eigenstates. We make our measurements using eigenstates of electric charge; these have strange mass ratios, e.g. the muon is 206 times heavier than the electron. If we were to make our measurements using eigenstates of square-root mass, we would find that all three generations have the mass ratios (1, 4, 9) whereas this time around the electric charge ratios will be strange. There is a perfect duality between electric charge and square-root mass.
A free electron in flight - is it in a charge eigenstate or a mass eigenstate? Neither! It is in a superposition of both, and collapses to one or the other, depending on what we choose to measure. In fact the free electron in flight does not separately have a mass and a charge; it has a quantum number which could be called charge -sqrt mass, which is the quantum number for the unified force. Unification is broken by measurement: if we measure EM effect then we attribue electric charge to the source. If we measure inertia or gravity, we attribute mass to the source. These statements are independent of energy scale. A classical measuring apparatus emerges from its quantum constituents as a consequence of sufficient entanglement: the emergence of such classical apparatus is the prelude to breaking of unification symmetry. In the early universe, sufficient entanglement is impossible above a certain energy [possibly the EW scale] and it appears as if symmetry breaking depends on energy. This is only an indirect dependence.  The true dependence of symmetry breaking is on the amount of entanglement. In our current low energy universe we have both low entanglement systems (quantum, unified) and high entanglement systems (classical, unification broken).

\noindent {\bf Some further insights into the origin of mass ratios for three fermion generations:} Prior to the Left-Right symmetry breaking, the $E_8 \times E_8$ symmetry is intact. We have three identical fermion generations of lepto-quark states  having an associated $U(1)$ quantum number which we call `electric-charge-square-root-mass'  and which takes the values $(0, 1/3, 2/3, 1)$. These lepto-quarks are excitations of the Dirac neutrino (quantum number 0) and can be labelled as LHdownquark-RHelectron (1/3), LHupquark-RHupquark (2/3), LHelectron-RHdownquark (1) with the understanding that all three generations are identical. The left-right symmetry breaking segregates these lepto-quark states into left chiral fermions with the associated $U(1)$ quantum number being electric charge, and the right chiral fermions with the associated $U(1)$ quantum number being square-root mass. The Dirac neutrino segregates into  a left-handed active Majorana neutrino and a right-handed sterile Majorana neutrino. The left-handed chiral fermions are excitations of the active left-handed Majorana neutrino, and are the anti-down quark (1/3), up quark (2/3) and positron (1) and their antiparticles. The right-handed chiral fermions are excitations of the sterile right-handed Majorana neutrino, and are the electron (1/3), up quark (2/3) and the down quark (1); numbers in bracket now show square-root mass. A Higgs coming from the right sector gives masses to the left-handed fermions; and remarkably, a charged Higgs coming from the left sector gives electric charge to the right-handed fermions. The gauge symmetry associated with the LH sector is the standard model symmetry $SU(3)_{color} \times SU(2)_L \times U(1)_Y$ and that associated with the RH sector is pre-gravitation $SU(3)_{grav} \times SU(2)_R \times U(1)_g$.

The left-handed states, being eigenstates of electric charge, are different from the right handed states which are eigenstates of square-root mass. When we solve the Dirac equation for three generations of a family with a given electric charge, assuming the neutrino to be Majorana,  it reveals mass quantisation, and the charge eigenstates are superpositions of square-root mass eigenstates and the corresponding eigenvalues carry information about the mass ratios across three generations. We called these eigenvalues Jordan eigenvalues \cite{Tp3}; they are shown in the table in Fig. 4.
\begin{figure}[h]
\centering
\includegraphics[width=15cm]{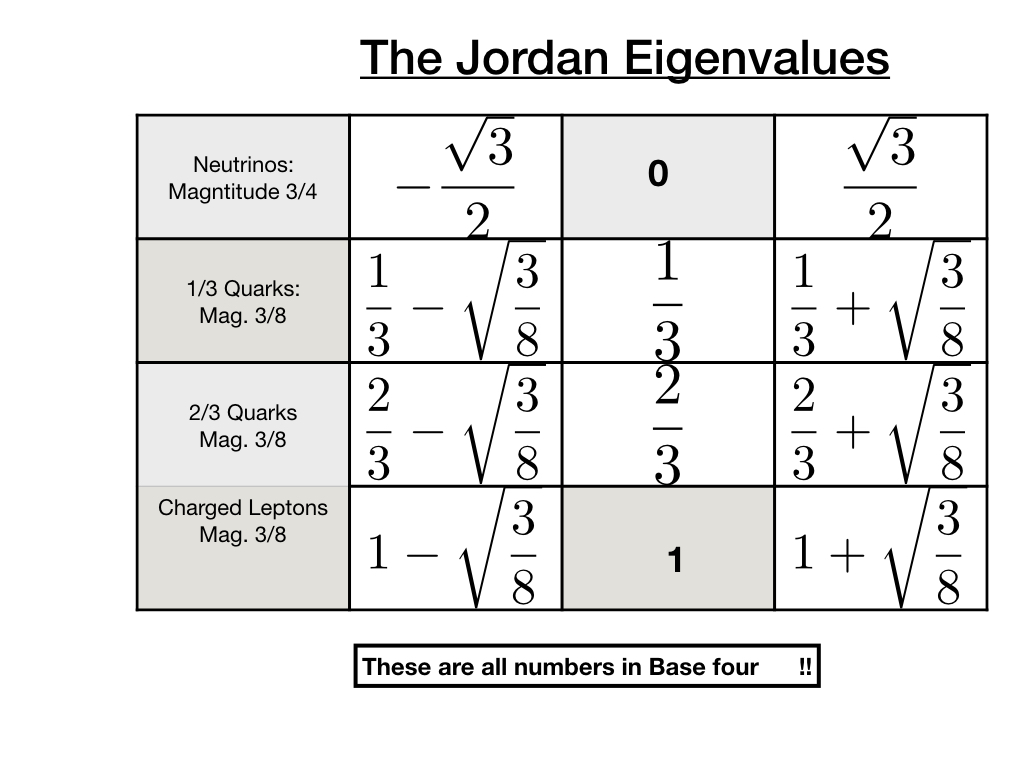}
\caption{The eigenvalues of the characteristic equation of the exceptional Jordan algebra, for quarks and leptons \cite{Tp3} }
\end{figure}
For a given electric charge value $q\neq 0$, the three eigenvalues take the form $q\pm \epsilon \sqrt{3/8}$ where $\epsilon=(-1,0,1)$. $\sqrt{3/8}$ is also the magnitude of the octonion which describes the state of the three generations of a family of charged fermions. For the neutrino, the $\sqrt{3/8}$ is replaced by $\sqrt{3/2}$. If the Jordan eigenvalues are calculated assuming the neutrino to be a Dirac particle then there is a subtle change in the values: the $\sqrt{3/8}$ is replaced by $\sqrt{3/2}$ for the charged leptons (no other change), and for the neutrino the eigenvalues are now very different: $(-1/2 -\sqrt{3/2}, 1, -1/2 + \sqrt{3/2})$. The shift from $\sqrt{3/2}$ to $\sqrt{3/8}$ results because  in going from the Dirac neutrino to the Majorana neutrino, the lepto-quark state splits into two halves - the LH charge eigenstate and the RH mass eigenstate - and hence the magnitude $\sqrt{3/2}$ is divided into two equal halves of $\sqrt{3/8}$ each. The Jordan eigenvalues hold for the anti-particles as well, with the charge $q$ being replaced by its negative value $-q$. Obviously, the eigenvalues of the neutrino do not change from the values shown in Fig. 4; consistent with the neutrino being its own anti-particle.
\begin{figure}[h]
\centering
\includegraphics[width=15cm]{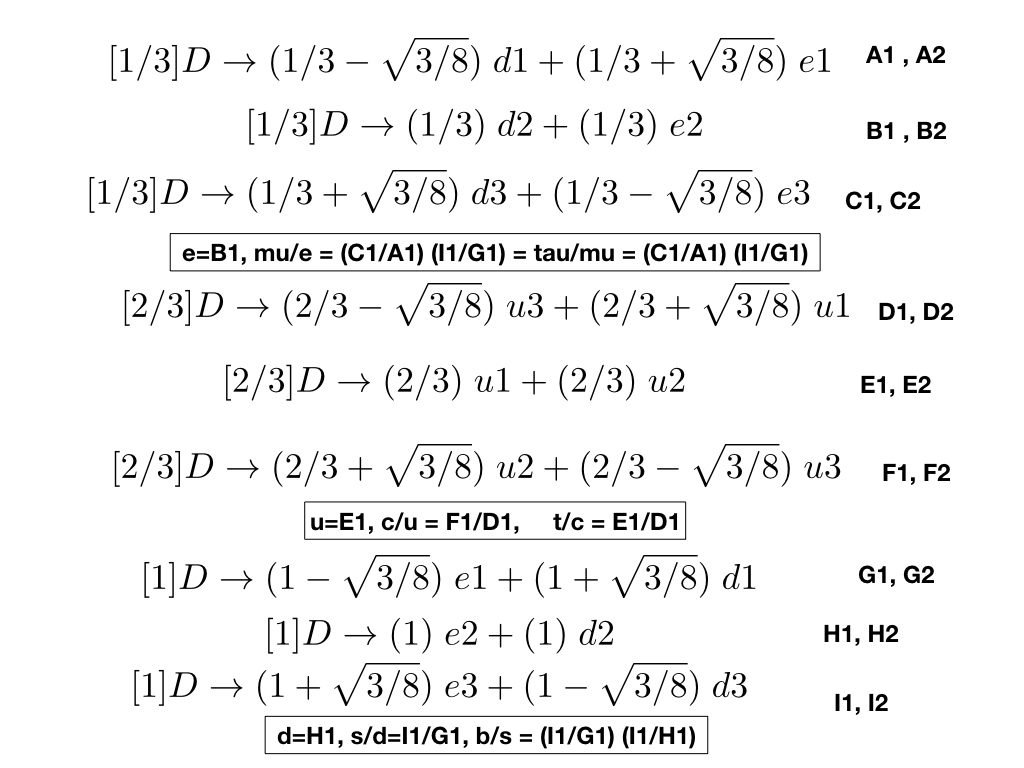}
\caption{From the Jordan eigenvalues, towards an understanding of the mass ratios. e1, e2, e3 denote the charged lepton family. d1, d2, d3 denote the down quark family. u1, u2, u3 denote the up quark family.  }
\end{figure}
Furthermore, these Jordan eigenvalues also represent superposition of square-root mass eigenstates in terms of charge eigenstates, with the down quark family interchanged with the charged lepton family. This is because the electric charge values $(1/3, 2/3, 1)$ respectively for the down quark family, up quark family, electron family are numerically also the same as square-root mass numbers respectively for the electron family, up quark family, down quark family. Thus the eigenvalues for the up quark family stay unchanged when charge and square-root mass number are interchanged for the electron family and down quark family. Because mass eigenstates are superpositions of charge eigenstates, the eigenvalues in the superposition determine the peculiar observed mass ratios, as we now elaborate in some detail, using the related data shown in Fig. 5

The first set of three rows are marked [1/3]D. D stands for `Dirac neutrino' and 1/3 is the `electric-charge-square-root-mass' quantum number prior to symmetry breaking, identical for three generations. The 4th, 5th, 6th rows are marked [2/3]D for the same quantum number taking the value 2/3. The 7th, 8th and 9th rows are marked [1]D because this quantum number takes the value 1. For a given value of this quantum number, the left-right symmetric lepto-quark   can be written as superposition of  a left handed fermion and a right handed fermion, with the  two Jordan eigenvalues in any of the nine rows giving the numerical coefficient of the superposition. The successive eigenvalues are labelled as the pairs (A1, B1), (A2, B2), (A3, B3), (A4, B4), (A5, B5), (A6, B6), (A7, B7), (A8, B8), (A9, B9). The first Jordan eigenvalue in any given row labels the left-handed fermion, and the second eigenvalue is the same row labels the right handed fermion.

When left-right symmetry breaking happens, the $U(1)$ quantum number prior to symmetry breaking becomes the $U(1)$ number electric charge for all LH fermions and takes the same set of values (0, 1/3,  2/3, 1) across the three generations. Analogously it takes the same set of values (0, 1/3, 2/3, 1) as the square-root mass number across the three generations of RH fermions, with the relative position of the electron and down quark interchanged. The up quark family positions stay unchanged at 2/3. Since we make all measurements using electric charge eigenstates (and not mass eigenstates), the mass eigenstates manifest in measurements as superpositions of electric charge eigenstates, and the ratios of Jordan eigenvalues reveal the observed mass ratios. Had we been making measurements using mass eigenstates, the electric charge eigenstates would manifest as superpositions of mass eigenstates, and we would have observed strange charge ratios.

Note that in every set of three rows for a given [1/n]D, there is a row where the left quantum number is the same as the right quantum number. These are the rows two, five and eight, with the numbers (B1, B2), (E1, E2) and (H1, H2). These give rise to the mass ratios (1/3, 2/3, 1) for the lightest charged fermions: electron, up quark, down quark. The other two rows with any given {1/n]D flip eigenvalues. These are the rows one and three, four and six, seven and nine. That is, (A1, A2) interchanges eigenvalues with (C1, C2), (D1, D2) with (F1, F2), and (G1, G2) with (I1, I2). 

The mass ratios are arrived at, from these eigenvalues, as follows, the simplest case being the up quark family [up, charm, top] with its quarks labelled u1, u2, u3. The up quark has the square-root mass ratio u=E1=2/3. The charm to up square-root mass ratio is c/u = F1/D1 = $(2/3+\sqrt{3/8})/(2/3 - \sqrt{3/8})$. The top to charm ratio is $t/c = E1/D1 = (2/3) / (2/3 - \sqrt{3/8})$.

The down quark family (down, strange, bottom)  and the electron family (electron, muon, tau lepton) are mixed, as can be seen  in the first three and  last three rows. For the [1/3]D rows, the first entry in every row is from the down quark family, and the second entry from the electron family. Whereas the roles are reversed in the [1]D rows. 

As per the last three rows, the down quark has the square-root mass ratio d=H1=1. The strange quark has the square-root mass ratio s/d = I1/G1 = $(1+\sqrt{3/8})/(1-\sqrt{3/8})$. The bottom to strange ratio is (I1/G1) (I1/H1). The origin of the peculiar factor I1/H1 remains to be understood.

From the first three rows, we see that the electron has the square-root mass ratio e=B1 = 1/3. The muon has the square-root mass ratio mu/e=(C1/A1)/(I1/G1) and this same ratio also holds for tau/mu.

These ratios above are the same as those shown earlier in Fig. 2 We believe that the octonionic theory provides a reasonably good understanding of the observed mass ratios of charged fermions. The mixing of the down quark family and the electron family is possibly the result of a gauge-gravity duality and of the fact that the three generations are not entirely independent of each other but related by the triality property of SO(8) \cite{Priyank}. Another remarkable feature we observed is that the eigenmatrices corresponding to the Jordan eigenvalues for the charged fermions always have the diagonal entry as 1/3, irrespective of whether the associated quantum number is 1/3 or 2/3 or 1. This seems to suggest that all charged fermions are made of three base states that all have an associated quantum number 1/3. The possible consequences of these observations are currently being investigated.

\noindent {\bf  Octonions and the Koide formula:} 
The Koide formula is the following observation  for the experimentally measured masses of charged leptons 
\begin{equation}
    \frac{m_e + m_{\mu} + m_{\tau}}{(\sqrt{m_e} + \sqrt{m_{\mu}} + \sqrt{m_{\tau}})^2} = 0.666661(7) \approx \frac{2}{3}
    \label{koide}
\end{equation}
That is, this ratio is close to (and a little less than) 2/3, but not exactly 2/3. Remarkably, the octonionic theory explains when the ratio is exactly 2/3, and why it departs from that exact value. 

Prior to the L-R symmetry breaking, we can consider that a left-handed-electron-right-handed-electron state has an associated electric-charge-square-rot-mass of 1, and the neutrino is a Dirac fermion. In this case, the Jordan eigenvalues, as we mentioned above, are $(1-\sqrt{3/2}, 1, 1+\sqrt{3/2})$. These give the superposition amplitudes when RH mass eigenstates are expressed as superposition of LH mass eigenstates. The Koide ratio is then
\begin{equation}
    \frac{(1+\sqrt{3/2})^2 + (1)^2 + (1-\sqrt{3/2})^2}{3^2} = \frac{2}{3}
    \label{koidedirac}
\end{equation}
Thus the exact value 2/3 is realised prior to symmetry breaking and prior to when the RH electron and RH down quark switch places. This switch might help understand why the mass ratios for charged leptons know about the Jordan eigenvalues $(1 + \sqrt{3/8})$ and $(1-\sqrt{3/8})$ which are otherwise associated with the down quark family.

Using our theoretical mass ratios for the charged leptons as reported in Fig.  2 we get the following theoretically predicted value for the Koide ratio
\begin{equation}
   \frac{m_e + m_{\mu} + m_{\tau}}{(\sqrt{m_e} + \sqrt{m_{\mu}} + \sqrt{m_{\tau}})^2}    = 0.669163 \approx \frac{2}{3}
\end{equation}
which is greater than the experimentally measured value of the Koide ratio and also greater than 2/3. The departure from the exact value of 2/3 is a consequence of the L-R symmetry breaking and of the switch between the RH down quark and RH electron. (It remains to be seen if the Karolyhazy correction will predict an exact match between theory and experiment).

Since unification already takes place at low energies (i.e. whenever the system is quantum and not yet measured upon) it follows that before we make a measurement on the charged leptons to measure their masses, the Koide ratio is exactly 2/3. After the measurement is made, the theoretical prediction for the resulting value is 0.669163, whereas the measured value is  smaller than $2/3$. The uncertainty in the mass of the tau-lepton 1776.86(12) MeV is such that by demanding the Koide ratio to be 2/3 one can predict the mass of the tau-lepton  to be 1776.969 MeV. At the upper limit 1776.98 MeV of the experimentally measured tau-lepton mass, the ratio is 0.66666728706, i.e. larger than 2/3, but smaller than our predicted theoretical value for Minkowski spacetime (the value realised after measurement). 

The above is an important result as we now know when the Koide ratio is exactly 2/3 [it is when the electron is not being observed]. And we understand why the measured value of this ratio is not exactly 2/3. In principle, we could have demanded the measured value to be equal to the theoretical value, and thereby fix the mass of the tau-lepton. It turns out there are no such allowed values for the tau-lepton mass, which is further evidence that the measured value could lie between the spinor spacetime value (2/3) and the Minkowski spacetime value if the mass of the tau lepton is greater than 1776.969 MeV. This is also indirect evidence that sterile neutrinos exist.

\noindent {\bf Why is matter electrically neutral?}
When L-R symmetry breaking mechanism in the early universe separated matter from anti-matter, particles were segregated from their anti-particles. And yet, the sign of the electric charge was not the criterion for deciding what went where. Matter has the positively charged up quark (2/3) and the negatively charged down quark (-1/3) and the electron (-1). Anti-matter has their anti-particles. If sign of electric charge was the deciding criterion for separating matter from anti-matter, all particles in our universe ought to have had the same sign of charge. That is not the case, and yet matter is electrically neutral! How could that have come about?
Even the algebraic proof based on the octonions, which shows quantisation of electric charge, naturally clubs positively charged particles together, when their states are made from a Clifford algebra:
\begin{align*}
& {\rm Particles} \quad Electric\ charge       \qquad     \qquad     \qquad       {\rm Anti-particles} \quad Electric \ charge \\
& Anti- Neutrino\   0              \qquad        \qquad \qquad        \qquad          Neutrino\ 0\\
& Antidown quark \ 1/3     \qquad        \qquad \qquad        \qquad                                                Down \  -1/3\\
& Up-quark\ 2/3               \qquad        \qquad \qquad        \qquad                     \qquad               Anti-up \ -2/3\\
& Positron \ 1        \qquad        \qquad \qquad        \qquad                         \qquad        \qquad                     Electron  \ -1
\end{align*} 
What picks the up quark from the left, and down and electron from the right, and clubs them as matter, and yet maintain electrical neutrality?
We have proposed that the criterion distinguishing matter from anti-matter is square-root of mass, not electric charge. One can make a new Clifford algebra afresh from the octonions, and show that square-root of mass is quantised, as in the above mass table.
Let us now calculate the net electric charge of matter, remembering that there are three down quarks (color) and three up quarks (color):
0 + (-1x1) + (3 x 2/3) + (3 x -1/3) =  0
It seems remarkable that the sum of the electric charges of matter (particles with +ve sqrt mass) comes out to be zero. It need not have been so. This demonstration might help understand how matter-antimatter separation preserved electrical neutrality.
Before this separation, the net square-root mass of matter and anti-matter was zero, even though individual sqrt masses were non-zero. In this we differ from the standard gauge-theoretic picture of EW symmetry breaking and mass acquisition. In EW, particles are massless before symmetry breaking, because a mass term in the Lagrangian breaks gauge invariance. However, for us sqrt mass is not zero before the symmetry breaking - its non-zero value was already set at the Planck scale (and cosmological expansion scaled down actual mass values while preserving mass ratios). Indeed it is rather peculiar if prior to the symmetry breaking particles have electric charge but no mass. For us, QFT on a spacetime background (and hence gauge theories) are not valid before the left-right symmetry breaking. In fact spacetime itself, along with gravitation, emerge after this symmetry breaking, as a result of the quantum to classical transition. Spacetime emerges iff classical matter emerges. 
Prior to the symmetry breaking, dynamics is described by trace dynamics, there is no spacetime, and we have `atoms' of space-time-matter. The concepts of electric charge and mass are not defined separately; there is only a charge-square-root mass [a hypercharge can also be defined, as for EW] and this is the source for a unified force in octonionic space.

\noindent {\bf Octonions, scale invariance, and a CPT symmetric universe: a possible explanation for the origin of matter-antimatter asymmetry}:
In the octonionic theory, prior to the so-called left-right symmetry breaking, the symmetry group is $E_8\times E_8$  and the Lagrangian of the theory is scale invariant. There is only one  parameter, a length scale, which appears as an overall multiplier of the trace Lagrangian.
Something dramatic happens after the symmetry breaking. Three new parameters emerge, to characterise the fermions:
Electric charge, has two signs, sign change operation C is complex conjugation, and ratios are (0, 1/3, 2/3, 1).
Chirality / spin, has two signs, sign change operation P is octonionic conjugation. Ratios (1/2, -1/2).
Square-root of mass, has two signs, sign change operation T is time reversal $t \rightarrow -t$ . Ratios are (0, 1/3, 2/3, 1).
Thus there are 2x2x2 = 8 types of fermions, based on sign of charge, square-root  mass, and spin.
This could possibly offer an attractive explanation for the origin of matter-antimatter asymmetry: a CPT symmetric universe. The four types of fermions which have positive square root mass become matter, our universe, moving forward in time. The other four types of fermions, which have negative square root mass, become anti-matter, a mirror universe moving backward in time! The forward moving universe and the backward moving universe together restore CPT symmetry. Our universe by itself violates T, and hence also CP. Matter and anti-matter repel each other gravitationally, thus explaining their separation. This also explains why gravitation in our universe is attractive, even though the underlying pre-gravitation theory is a vector interaction.
Prior to the symmetry breaking, an octonionic inflation [scale invariant, time-dependent in Connes time] precedes the `big bang' creation event, which is the  symmetry breaking itself. Freeze out happens when radiation $\rightarrow$ matter-antimatter is no longer favorable. Segregation takes place; our matter universe has a one in a billion excess of matter over anti-matter. The backward in time mirror universe has a one in a billion excess of anti-matter over matter.
The maths of complex octonions naturally accounts for the C, P, T operations. Scale invariance is transformed into CPT invariance in the emergent universe. We hope to  make this idea rigorous in forthcoming investigations.
In an elegant proposal, Turok and Boyle \cite{Turok} have also recently proposed a CPT symmetric universe [mirror universes]. They, however, did not use the octonions.

\noindent {\bf Prospects of tests through particle physics experiments and phenomenology:} 
In Section XV of \cite{Singhreview} we have briefly discussed some of the possible experimental predictions of this theory and prospects for their experimental tests. Below we discuss some particle physics related predictions and their possible relevance for collider experiments and neutrino experiments. These ideas are largely based on a detailed conversation with Ashutosh Kotwal, and their compilation by Vatsalya Vaibhav. To them I express my gratitude, for stimulating discussions.

Our theory predicts specific new particles, though much work remains to be done to predict their specific properties such as masses. We predict that the neutrino is Majorana and that there are three right-handed sterile neutrinos. At present we do not understand neutrino masses, though there is a very real possibility that the neutrino is massless and flavor oscillations are caused by spacetime being higher dimensional and octonionic, and the neutrinos being spacetime triplets \cite{Priyank}. This possibility could br verified if we can calculate the PMNS matrix from first principles in the octonionic theory; this will be attempted in future work. The Majorana nature of the neutrino also suggests neutrinoless double beta decay, experimental implications for which we will investigate further. The possibility that the sterile neutrino could be massive and light (hence hot dark matter) or heavy (hence cold dark matter) will also be investigated, although it is true that our theory favours MOND over cold dark matter.

Our theory also predicts a new charged Higgs boson, and possibly a doubly charged Higgs, although their masses and the fundamental origin of their scalar nature remains to be understood. Also of interest is whether the Higgs triplet in this theory can cause the W mass to depart from the standard model prediction. The hierarchy problem, and the exact mechanism of the electroweak symmetry breaking and the left-right symmetry breaking remain to be understood: these could be the same symmetry breaking and could perhaps be mediated by a quantum-to-classical phase transition. It is also of interest to try and predict the (g-2) anomaly, some related physics was discussed in \cite{MPSingh} and this issue might also be related to our derivations of mass ratios and the low energy fine structure constant.

A general line of investigation of serious interest is to note that our predictions are made on octonionic twistor space, which is non-classical, whereas all measurements are in Minkowski spacetime, and based on quantum field theory calculations. In transiting from trace dynamics to emergent quantum field theory there could be important and smoking gun corrections, which could possibly be cast in the language of effective field theory.

As per our recent work on $E_8 \times E_8$ symmetry, we have been able to account for 208 out of the 496 degrees of freedom \cite{Priyank}. While we have commented briefly in that work, on the unaccounted degree of freedom, more work needs to be done to see if they predict new particles which can confirm or rule out this theory.

\noindent {\bf Why there is no black hole information loss paradox in the octonionic theory?}

There appears to be an information loss paradox because we are ignoring the physical process of the quantum-to-classical transition which keeps the black hole (made of enormously many quantum constituents) classical in the first place.
Let us consider the following analogy:
Consider a box of gas at thermodynamic equilibrium - this is the maximum entropy and minimum information state. Now, a sudden spontaneous fluctuation sends the entire set of gas molecules to one corner of the box. This is a transient low entropy high information and ordered state, far from thermodynamic equilibrium. At the very next instant, the gas molecules will return to equilibrium, spread all over the box, entropy will have been gained, and information lost. If we ignore the spontaneous fluctuation which sent the gas to a corner, then we have an information loss paradox! Obviously there is no paradox in reality: information gained during the spontaneous fluctuation is lost during the return to equilibrium.
It is exactly the same physics, when we work with a fundamental theory (generalised trace dynamics valid at Planck time resolution) from which quantum theory and classical gravitation are emergent phenomena. Quantum theory (without classical time)  is emergent as the thermodynamic equilibrium state description in a statistical thermodynamic approximation of the underlying theory. Classical gravitation, spacetime, and hence the black hole, is a far from equilibrium transient state which arises from a spontaneous fluctuation caused by non-unitary evolution. At equilibrium, the evolution is unitary, except when a large fluctuation kicks in.
In Adler's theory of trace dynamics, a collection of quantum states at thermodynamic equilibrium constitute a state of maximum entropy (the most likely state to emerge from coarse-graining of the underlying system).  Departure from equilibrium by way of a spontaneous fluctuation produces the black hole as a state of relatively low entropy and high information. By way of Hawking evaporation, the black hole returns to the state of maximum entropy and thermodynamic equilibrium. [One has to be careful here while talking of pure states and mixed states; there is no classical  spacetime at thermodynamic equilibrium, in this theory].
To this, one could  object that not all black holes are formed in the way suggested in the previous paragraph above. That is true; if one has in mind stellar gravitational collapse ending in black hole formation. However, let us trace the star all the way back to the very early universe: inflationary quantum perturbations  became classical, spacetime emerged, density perturbations grew, star  formed, black hole formed, it Hawking radiated, and is going back to the inflationary stage which was quantum in nature. 
The formation of classical large scale structure in the universe is like the gas in the box spontaneously fluctuating away from equilibrium. Hawking evaporation is like the gas returning to equilibrium.
Interestingly, the above reasoning also throws light on the current accelerating phase of the universe. de Sitter like inflationary expansion is the natural expansion state of the universe - the equilibrium quantum state. Emergence of classicality gives rise to the radiation dominated and matter dominated eras, where matter density is high enough to dictate a power-law expansion of the scale factor, overriding de Sitter like expansion. In recent cosmic history, matter density is falling, and hence de Sitter like scale-invariant expansion [quantum equilibrium] is again dominating. This dynamics is very likely related to the emergent classical Lagrangian of our theory, which exhibits modification of general relativity by conformal gravity, and will be investigated further for its cosmological implications.

\begin{figure}[h]
\centering
\includegraphics[width=15cm]{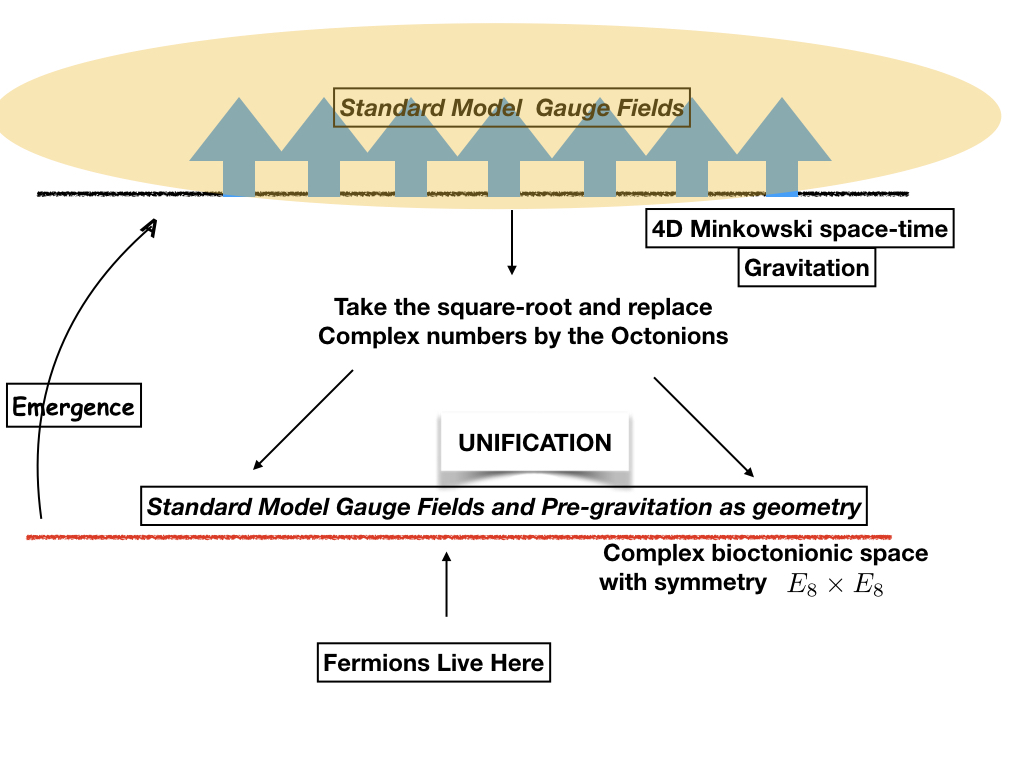}
\caption{Taking the square-root of 4D Minkowski space-time paves the way to unification}
\end{figure}

\noindent {\bf How taking the square-root of Minkowski space-time paves the way for unification:}
In summary, we believe we have a promising theory of unification under development, as captured in Fig. 6 and explained briefly below.

It is like going from the surface of the ocean to the ocean bed. The ocean floor can exist without the surface, but the surface cannot exist without the floor.
We live in a 4D Minkowski space-time curved by gravitation, in which standard model gauge fields and fermions reside. 
But there is a more precise description. We take the square-root of Minkowski space-time and arrive at Penrose's twistor space, described by complex numbers.
In this spinor space-time replace complex numbers by quaternions, then by octonions. More precisely complex split bioctonions. We arrive at a space with $E_8 \times E_8$ symmetry whose geometry is a unified description of the standard model and pre-gravitation. The gauge group is 
$SU(3)_c \times SU(2)_L \times U(1)_Y \times SU(3)_{grav} \times SU(2)_R \times U(1)_g$
Coupling constants are determined by the geometry.
In the classical limit, the 4D curved spacetime and the standard model emerge, but with pre-determined values of coupling constants. Fermions span space-time as well as the space of the gauge fields.
Taking the square root of Minkowski space-time does not involve change of energy scale. It only gives a more precise mathematical  formalism. One that is key to unification, which already takes place at low energies, if we do not restrict ourselves to 4D classical spacetime: only classical systems live in 4D. Quantum systems always live in $E_8 \times E_8$ even at low energies. If we want a comparison with string theory, then this new theory is string theory without compactification. Compactification is effectively achieved by the quantum-to-classical transition; it does not have to be enforced in an ad hoc manner.

\bigskip

\bigskip

\noindent {\bf Acknowledgements:} It is a pleasure to thank Priyank Kaushik, Ashutosh Kotwal, Sherry Raj and Vatsalya Vaibhav for very helpful discussions.

\newpage


\end{document}